\documentclass[zpreprint,zbstdefault]{./LaTeX/zeus/zeus_paper}
\usepackage[english]{babel}

\newcommand{\ZcoosysB}{%
The ZEUS coordinate system is a right-handed Cartesian system, with the $Z$
axis pointing in the proton beam direction, referred to as the ``forward
direction'', and the $X$ axis pointing left towards the centre of HERA.
The coordinate origin is at the nominal interaction point.\xspace}
\newcommand{\Zpsrap}{%
The pseudorapidity is defined as $\eta=-\ln\left(\tan\frac{\theta}{2}\right)$,
where the polar angle, $\theta$, is measured with respect to the proton beam
direction.\xspace}

\newcommand{\ZcoosysfnBeta}{\footnote{\ZcoosysB\Zpsrap}}
\newcommand{\Zdetdesc}{%
A detailed description of the ZEUS detector can be found 
elsewhere~\cite{zeus:1993:bluebook}. A brief outline of the 
components that are most relevant for this analysis is given
below.\xspace}
\newcommand{\Zctddesc}[1]{%
Charged particles were tracked in the central tracking detector (CTD)~\citeCTD,
which operated in a magnetic field of $1.43\Tesla$ provided by a thin 
superconducting coil. The CTD consisted of 72~cylindrical drift chamber 
layers, organised in 9~superlayers covering the polar-angle#1 region 
\mbox{$15^\circ<\theta<164^\circ$}. The transverse-momentum resolution for
full-length tracks was $\sigma(p_T)/p_T=0.0058p_T\oplus0.0065\oplus0.0014/p_T$,
with $p_T$ in $\Gev$.}
\newcommand{\Zcaldesc}{%
The high-resolution uranium--scintillator calorimeter (CAL)~\citeCAL consisted  
of three parts: the forward (FCAL), the barrel (BCAL) and the rear (RCAL)
calorimeters. Each part was subdivided transversely into towers and
longitudinally into one electromagnetic section (EMC) and either one (in RCAL)
or two (in BCAL and FCAL) hadronic sections (HAC). The smallest subdivision of
the calorimeter was called a cell.  The CAL energy resolutions, as measured under
test-beam conditions, were $\sigma(E)/E=0.18/\sqrt{E}$ for electrons and
$\sigma(E)/E=0.35/\sqrt{E}$ for hadrons, with $E$ in $\Gev$.}

\chardef\usc=95
\chardef\til=126
\catcode`\@=11 
\DeclareRobustCommand\xdotspace{\futurelet\@let@token\@xdotspace}
\def\@xdotspace{%
  \ifx\@let@token.\else
  \ifx\@let@token\bgroup.\else
  \ifx\@let@token\egroup.\else
  \ifx\@let@token\/.\else
  \ifx\@let@token\ .\else
  \ifx\@let@token~.\else
  \ifx\@let@token!.\else
  \ifx\@let@token,.\else
  \ifx\@let@token:.\else
  \ifx\@let@token;.\else
  \ifx\@let@token?.\else
  \ifx\@let@token/.\else
  \ifx\@let@token'.\else
  \ifx\@let@token).\else
  \ifx\@let@token-.\else
  \ifx\@let@token\@xobeysp.\else
  \ifx\@let@token\space.\else
  \ifx\@let@token\@sptoken.\else
   .\space
   \fi\fi\fi\fi\fi\fi\fi\fi\fi\fi\fi\fi\fi\fi\fi\fi\fi\fi}
\catcode`\@=12 

\newcommand{\stru}[2]{%
   \relax\ifmmode\hbox{\vrule height#1 depth#2 width0pt}%
   \else\vrule height#1 depth#2 width0pt\fi}

\newcommand{\Ronum}[1]{\uppercase\expandafter{\romannumeral#1}}
\newcommand{\ronum}[1]{\expandafter{\romannumeral#1}}
\DeclareRobustCommand{\LaTeXZ}{%
  \LaTeX\kern-.05em4\kern-.1em
  {\raisebox{-0.2ex}{$\scriptstyle\text{ZEUS}$}}\xspace}

\DeclareMathAlphabet{\mathbf}{OT1}{cmr}{bx}{sl}
\newcommand{\eVdist}{\kern-0.06667em}

\newcommand{\Gev}{{\text{Ge}\eVdist\text{V\/}}}

\newcommand{\gev}{{\,\text{Ge}\eVdist\text{V\/}}}

\newcommand{\Tesla}{\,\text{T}}

\newcommand{\slashfrac}[2]{%
  \raisebox{0.5ex}{\ensuremath #1}\kern-0.12em/\kern-0.08em
  \raisebox{-.8ex}{\ensuremath #2}}

\newcommand{\sqr}[3]{%
    {\vcenter{\hrule height.#3ex\hbox{\vrule width.#2ex height#1ex
     \kern#1ex\vrule width.#3ex}\hrule height.#2ex}}}

\catcode`\@=11 
\newcommand{\parenbar}{\mathpalette\p@renb@r}
\def\p@renb@r#1#2{\vbox{%
  \ifx#1\scriptscriptstyle \dimen@.7em\dimen@ii.2em\else
  \ifx#1\scriptstyle \dimen@.8em\dimen@ii.25em\else
  \dimen@1em\dimen@ii.4em\fi\fi \offinterlineskip
  \ialign{\hfill##\hfill\cr
    \vbox{\hrule width\dimen@ii}\cr
    \noalign{\vskip-.3ex}%
    \hbox to\dimen@{$\mathchar300\hfil\mathchar301$}\cr
    \noalign{\vskip-.3ex}%
    $#1#2$\cr}}}
\catcode`\@=12 

\newcommand{\IP}{{\rm I$\kern-0.01667em$P}\xspace}

\mathchardef\qsm=63
\mathchardef\pls=43
\mathchardef\mns=512
\mathchardef\plm=518
\mathchardef\eql=61
\mathchardef\smallleft=300
\mathchardef\smallright=301
\mathchardef\les=316
\mathchardef\gre=318
\mathchardef\leq=532
\mathchardef\grq=533

\catcode`\@=11 
\newcounter{pict@width}
\newcounter{pict@height}
\newlength{\pict@scale}
\newcommand{\psfigadd}[4]{%
\setcounter{pict@width}{1*\ratio{#2+\pict@scale/2}{\pict@scale}}
\setcounter{pict@height}{1*\ratio{#3+\pict@scale/2}{\pict@scale}}
\setlength{\unitlength}{\pict@scale}
\hbox to #2{\hspace{-\fill}\begin{picture}(\thepict@width,\thepict@height)
\put(0,0){\psfig{figure=#1,width=#2,height=#3,clip=}}
\SetScale{0.283466457}
\SetWidth{1.763889}
{#4}
\end{picture}}
}
\newcounter{pict@widthfst}
\newcounter{pict@widthscd}
\newcounter{pict@widthtot}
\newcommand{\psfigaddtwo}[7]{%
\setcounter{pict@widthfst}{1*\ratio{#2+\pict@scale/2}{\pict@scale}}
\setcounter{pict@widthscd}{1*\ratio{#2+#4+\pict@scale/2}{\pict@scale}}
\setcounter{pict@widthtot}{1*\ratio{#2+#4+#6+\pict@scale/2}{\pict@scale}}
\setcounter{pict@height}{1*\ratio{#3+\pict@scale/2}{\pict@scale}}
\setlength{\unitlength}{\pict@scale}
\hbox{\hspace{-\fill}\begin{picture}(\thepict@widthtot,\thepict@height)
\put(0,0){\psfig{figure=#1,width=#2,height=#3,clip=}}
\put(\thepict@widthscd,0){\psfig{figure=#5,width=#6,height=#3,clip=}}
\SetScale{0.283466457}
\SetWidth{1.763889}
{#7}
\end{picture}}
}
\newcommand{\psfigror}[4]{%
\setcounter{pict@width}{1*\ratio{#2+\pict@scale/2}{\pict@scale}}
\setcounter{pict@height}{1*\ratio{#3+\pict@scale/2}{\pict@scale}}
\setlength{\unitlength}{\pict@scale}
\hbox{\begin{picture}(\thepict@width,\thepict@height)
\put(0,\thepict@height){\psfig{figure=#1,width=#3,height=#2,clip=,angle=270}}
\SetScale{0.283466457}
\SetWidth{1.763889}
{#4}
\end{picture}}
}
\newcommand{\psfigrol}[4]{%
\setcounter{pict@width}{1*\ratio{#2+\pict@scale/2}{\pict@scale}}
\setcounter{pict@height}{1*\ratio{#3+\pict@scale/2}{\pict@scale}}
\setlength{\unitlength}{\pict@scale}
\hbox{\begin{picture}(\thepict@width,\thepict@height)
\put(0,0){\psfig{figure=#1,width=#3,height=#2,clip=,angle=90}}
\SetScale{0.283466457}
\SetWidth{1.763889}
{#4}
\end{picture}}
}
\catcode`\@=12 
\newlength\listtextwidth

\catcode`\@=11 
\newlength{\@tabfninsert}
\newlength{\@tabfnwidth}
\newcommand{\tabfootnote}[2]{%
  \setlength{\@tabfninsert}{0.8em}
  \setlength{\@tabfnwidth}{\textwidth}
  \addtolength{\@tabfnwidth}{-\@tabfninsert}
  \addtolength{\@tabfnwidth}{-0.4em}
  \noindent\makebox[\@tabfninsert][r]{\footnotesize$^{#1}$\hfil}\hfill%
  \parbox[t]{\@tabfnwidth}{\footnotesize #2\hfill}}
\catcode`\@=12 

\def\citeCTD{{\cite{%
nim:a279:290,*npps:b32:181,*nim:a338:254%
}}\xspace}
\def\citeCAL{{\cite{%
nim:a309:77,*nim:a309:101,*nim:a321:356,*nim:a336:23%
}}\xspace}

\begin{document}
%
\title{
\vspace{-5cm}
\begin{flushright} {\normalsize \tt DESY 08-209}\\ \vspace{-.25cm}{\normalsize 
\tt December 2008} \end{flushright}
\vspace{2cm}
Measurement of the charm fragmentation \\
function in \boldmath{$D^*$} photoproduction at HERA
}
                    
\author{ZEUS Collaboration}
\draftversion{Post-reading}
\date{\today}

\abstract{
The charm fragmentation function has been measured in $D^*$ photoproduction with the 
ZEUS detector at HERA using an integrated luminosity of 120\,pb$^{-1}$. The 
fragmentation function is measured versus $z = (E+p_\parallel)^{D^*}/2E^{\rm jet}$,  
where $E$ is the energy of the $D^*$ meson and $p_\parallel$ is the longitudinal 
momentum of the $D^*$ meson relative to the axis of the associated jet of energy 
$E^{\rm jet}$. Jets were reconstructed using the 
$k_T$ clustering algorithm and required to have transverse energy larger than 9\,GeV. 
The $D^*$ meson associated with the jet was required to have a transverse momentum 
larger than 2\,GeV. 
The measured function is compared to different fragmentation 
models incorporated in leading-logarithm 
Monte Carlo simulations and in a next-to-leading-order QCD calculation.  The free 
parameters in each fragmentation model are fitted to the data. 
The extracted 
parameters and the function itself are compared to measurements from $e^+e^-$ 
experiments.
}

\makezeustitle

\def\3{\ss}                                                                                        
\pagenumbering{Roman}                                                                              
                                                   %
\begin{center}                                                                                     
{                      \Large  The ZEUS Collaboration              }                               
\end{center}                                                                                       
  S.~Chekanov,                                                                                     
  M.~Derrick,                                                                                      
  S.~Magill,                                                                                       
  B.~Musgrave,                                                                                     
  D.~Nicholass$^{   1}$,                                                                           
  \mbox{J.~Repond},                                                                                
  R.~Yoshida\\                                                                                     
 {\it Argonne National Laboratory, Argonne, Illinois 60439-4815, USA}~$^{n}$                       
\par \filbreak                                                                                     
  M.C.K.~Mattingly \\                                                                              
 {\it Andrews University, Berrien Springs, Michigan 49104-0380, USA}                               
\par \filbreak                                                                                     
  P.~Antonioli,                                                                                    
  G.~Bari,                                                                                         
  L.~Bellagamba,                                                                                   
  D.~Boscherini,                                                                                   
  A.~Bruni,                                                                                        
  G.~Bruni,                                                                                        
  F.~Cindolo,                                                                                      
  M.~Corradi,                                                                                      
\mbox{G.~Iacobucci},                                                                               
  A.~Margotti,                                                                                     
  R.~Nania,                                                                                        
  A.~Polini\\                                                                                      
  {\it INFN Bologna, Bologna, Italy}~$^{e}$
% ...
\newpage                                                                                           
\enlargethispage{5cm}                                                                              
$^{\    1}$ also affiliated with University College London,                                        
United Kingdom\\                                                                                   
% ...
$^{\    5}$ supported by the research grant no. 1 P03B 04529 (2005-2008) \\                        
% ...
%
\newpage   
                                                           %
                                                           %
\begin{tabular}[h]{rp{14cm}}                                                                       
$^{a}$ &  supported by the Natural Sciences and Engineering Research Council of Canada (NSERC) \\  
$^{b}$ &  supported by the German Federal Ministry for Education and Research (BMBF), under        
          contract numbers 05 HZ6PDA, 05 HZ6GUA, 05 HZ6VFA and 05 HZ4KHA\\                         
$^{c}$ &  supported in part by the MINERVA Gesellschaft f\"ur Forschung GmbH, the Israel Science   
          Foundation (grant no. 293/02-11.2) and the U.S.-Israel Binational Science Foundation \\  
$^{d}$ &  supported by the Israel Science Foundation\\                                             
$^{e}$ &  supported by the Italian National Institute for Nuclear Physics (INFN) \\                
$^{f}$ &  supported by the Japanese Ministry of Education, Culture, Sports, Science and Technology 
          (MEXT) and its grants for Scientific Research\\                                          
$^{g}$ &  supported by the Korean Ministry of Education and Korea Science and Engineering          
          Foundation\\                                                                             
$^{h}$ &  supported by the Netherlands Foundation for Research on Matter (FOM)\\                   
$^{i}$ &  supported by the Polish State Committee for Scientific Research, project no.             
          DESY/256/2006 - 154/DES/2006/03\\                                                        
$^{j}$ &  partially supported by the German Federal Ministry for Education and Research (BMBF)\\   
$^{k}$ &  supported by RF Presidential grant N 1456.2008.2 for the leading                         
          scientific schools and by the Russian Ministry of Education and Science through its      
          grant for Scientific Research on High Energy Physics\\                                   
$^{l}$ &  supported by the Spanish Ministry of Education and Science through funds provided by     
          CICYT\\                                                                                  
$^{m}$ &  supported by the Science and Technology Facilities Council, UK\\                         
$^{n}$ &  supported by the US Department of Energy\\                                               
$^{o}$ &  supported by the US National Science Foundation. Any opinion,                            
findings and conclusions or recommendations expressed in this material                             
are those of the authors and do not necessarily reflect the views of the                           
National Science Foundation.\\                                                                     
$^{p}$ &  supported by the Polish Ministry of Science and Higher Education                         
as a scientific project (2006-2008)\\                                                              
$^{q}$ &  supported by FNRS and its associated funds (IISN and FRIA) and by an Inter-University    
          Attraction Poles Programme subsidised by the Belgian Federal Science Policy Office\\     
$^{r}$ &  supported by an FRGS grant from the Malaysian government\\                               
\end{tabular}                                                                                      
                                                           %
                                                           %

\pagenumbering{arabic} 
\pagestyle{plain}
\section{Introduction}
\label{sec-int}

The production of a charm hadron is described as the convolution of the
perturbative production of a charm quark and the non-perturbative transition
of a charm quark to a hadron.  The non-perturbative component is assumed to be 
universal, i.e.\ independent of the initial conditions.  It is described by 
so-called fragmentation functions which parametrise the transfer of the quark's 
energy to a given hadron. The free parameters are determined from fits to data.
The transition of a charm quark to a $D^*$ meson is the subject of this paper. 

The parameters of the various fragmentation function ans\"{a}tze were so far derived from 
data obtained at $e^+e^-$ colliders. The $e^+e^-$ 
data span a wide range of centre-of-mass energies and the fragmentation of a charm 
quark to a $D^*$ meson has been measured many times\,\cite{proc:hera-lhc:2005:b390}, 
most recently by the CLEO\,\cite{pr:d70:112001} and Belle\,\cite{pr:d73:032002} 
collaborations at a centre-of-mass energy of $\sim$10.5\,GeV and the 
ALEPH\,\cite{epj:c16:597} collaboration at 91.2\,GeV. Due to scaling violations in QCD, 
the dependence of the fragmentation function on production 
energy\,\cite{np:b361:626,proc:hera-lhc:2005:b390} is expected to follow the DGLAP 
equations\,\cite{sovjnp:15:438,*sovjnp:20:94,*np:b126:298,*jetp:46:641}.

The fragmentation function has recently been measured by the H1 Collaboration for 
the production of $D^*$ mesons in deep inelastic scattering (DIS)\,\cite{epj:c59:589}. 
A measurement of the fragmentation function at HERA and its comparison 
with that deduced from experiments at $e^+e^-$ colliders provides a measure of the 
universality of charm fragmentation and further constrains its form. The analysis 
presented here has been performed in the photoproduction regime in which a quasi-real 
photon of low virtuality, $Q^2$, is emitted from the incoming electron or positron and 
collides with a parton in the proton. 

\section{Experimental conditions}
\label{sec-method}
The analysis was performed using data collected with the ZEUS detector at HERA during 
1996--2000. In this period, HERA collided electrons or positrons with energy 
\mbox{$E_e\,=\,27.5\gev$} and protons with energy $E_p\,=\,820\gev$ (1996--1997) or 
$E_p\,=\,920\gev$ (1998--2000)  corresponding to integrated luminosities 
of $38.6\,\pm\,0.6$ and $81.9\,\pm\,1.8\,{\rm pb^{-1}}$ and to centre-of-mass energies 
$\sqrt{s}\,=\,300$\,GeV and $\sqrt{s}\,=\,318$\,GeV, respectively.
 
\Zdetdesc
 
\Zctddesc\ZcoosysfnBeta
 
\Zcaldesc
 
The luminosity was measured from the rate of the bremsstrahlung process
$ep~\rightarrow~e\gamma p$, where the photon was measured in a lead--scintillator
calorimeter\,\cite{desy-92-066,*zfp:c63:391,*acpp:b32:2025} placed in the HERA tunnel at 
$Z=-107~{\rm m}$.

\section{Event selection and reconstruction}
\label{sec-data}

A three-level trigger system was used to select events 
online\,\cite{zeus:1993:bluebook,uproc:chep:1992:222,epj:c1:109}. At the first- and second-level triggers, 
general characteristics of photoproduction events were required and background due to 
beam-gas interactions rejected. At the third level, a version of the tracking 
information close to the offline version was used to select $D^*$ candidates.

Kinematic variables and jets were reconstructed offline using a combination of track 
and calorimeter information that optimises 
the resolution of reconstructed kinematic variables\,\cite{epj:c6:43,*briskin:phd:1998}. 
A selected track or calorimeter cluster is referred to as an Energy Flow Object (EFO). 
The jets were reconstructed with the $k_T$ cluster algorithm\,\cite{np:b406:187} in its 
longitudinally invariant inclusive mode\,\cite{pr:d48:3160}, where the parameter $R$ is
 chosen equal to 1. Jets were formed from the 
EFOs with at least one jet required to have transverse energy, \mbox{$E_T^{\rm jet}>$ 9\,GeV} 
and pseudorapidity, 
\mbox{$|\eta^{\rm jet}|<$ 2.4}. The photon-proton centre-of-mass energy, $W_{\gamma p}$, 
was calculated using the formula $W_{\gamma p}\,=\,\sqrt{2 E_p (\sum_i E_i - p_{Z,i})}$, where 
the sum runs over the energy and longitudinal momentum component of all EFOs. Due to 
trigger requirements and beam-gas background at low $W_{\gamma p}$ and background from 
DIS events at high $W_{\gamma p}$, the requirement 
\mbox{$130 < W_{\gamma p} < 280$\,GeV} was made. Neutral current DIS events with a scattered 
electron or positron candidate in the CAL were also removed by cutting\,\cite{pl:b322:287} on the 
inelasticity, $y$, which is estimated from the energy, $E_e^\prime$, and polar angle, 
$\theta_e^\prime$, of the scattered electron or positron candidate using 
\mbox{$y_e=1-\frac{E_e^\prime}{2E_e}(1-\cos\theta_e^\prime)$}. Events were rejected if 
$y_e < 0.7$.

The $D^*$ mesons were identified using the decay channel $D^{*+} \to D^0 \pi_s^+$ with 
the subsequent decay $D^0 \rightarrow K^- \pi^+$ and the corresponding anti-particle 
decay. They were reconstructed from charged tracks in the CTD using 
the mass-difference technique\,\cite{prl:35:1672,*prl:38:1313}. Tracks with opposite charges 
and transverse momenta greater than 0.5\,GeV were combined into pairs to form $D^0$ 
candidates. No particle identification was used, so kaon and pion masses were assumed in 
turn for each track to calculate the invariant mass $M(K\pi)$. A third track, assumed to 
be the soft pion, $\pi_s^+$, with transverse momentum greater than 0.12\,GeV and of opposite 
charge to the kaon, was combined to form a $D^*$ candidate with invariant mass $M(K\pi\pi_s)$. The $D^*$ 
candidates were then required to have \mbox{$p_T^{D^*} > 2$\,GeV} and \mbox{$|\eta^{D^*}|<1.5$}. 

\begin{figure}[htb]
\begin{center}
~\epsfig{file=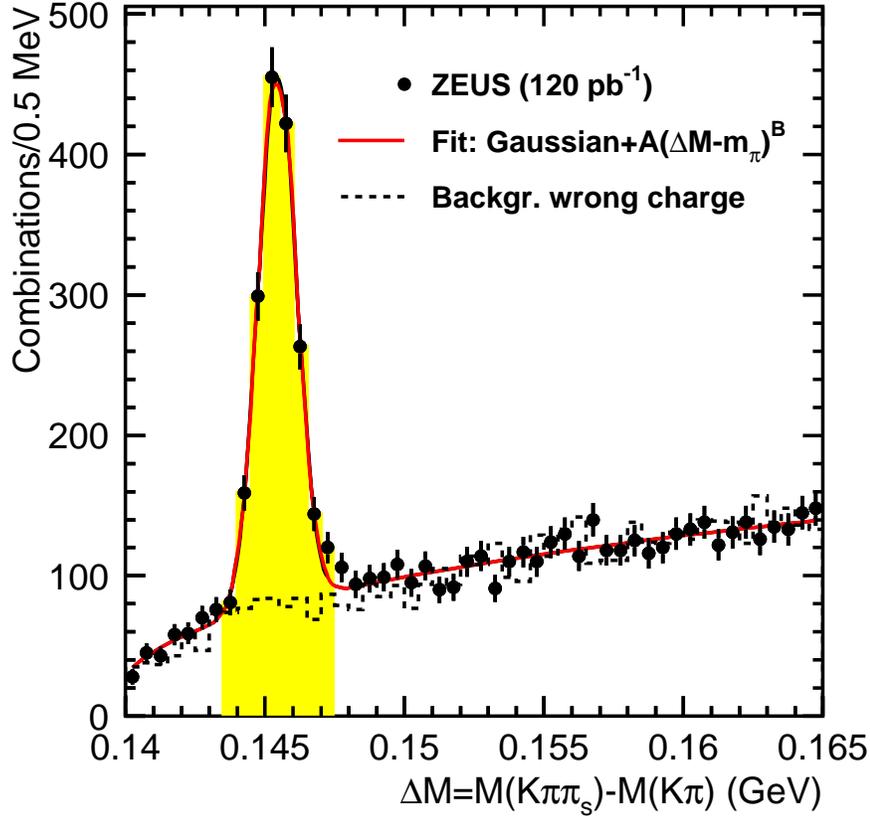,height=12cm}
\end{center}
\caption{The distribution of the mass difference, $\Delta M$, showing the right-charge 
combinations (points) and 
wrong-charge combinations (dashed histogram). The shaded area shows the signal region, 
\mbox{$0.1435 < \Delta M < 0.1475$\,GeV}. The solid line is a fit to a Gaussian function plus 
$A(\Delta M - m_\pi)^B$ to describe the background, where $A$ and $B$ are constants.}

\label{fig:mass}
\end{figure}

To minimise background, narrow windows were selected for the mass difference, 
$\Delta M = M(K\pi\pi_s) -M(K\pi)$, and the mass of the 
$D^0$ meson: \mbox{$0.1435 < \Delta M < 0.1475$\,GeV} and \mbox{$1.83 < M(D^0) < 1.90$\,GeV}. For 
background determination, $D^0$ candidates with wrong-charge combinations, in which 
both tracks forming the $D^0$ candidates have the same charge and the third track has the 
opposite charge, were also retained. The same kinematic restrictions were applied as for 
those $D^0$ candidates with correct-charge combinations. The normalisation factor of 
the wrong-charge sample (a value of 1.02 for the distribution after all requirements shown in 
Fig.~\ref{fig:mass}) was determined as the ratio of events with correct-charge combinations 
to wrong-charge combinations in the region \mbox{$0.150 < \Delta M < 0.165$\,GeV}. A cut of 
$p_T^{D^*}/E_\perp^{\theta>10^\circ} > $ 0.1 was imposed to further reduce combinatorial 
background, where $E_\perp^{\theta>10^\circ}$ is the transverse energy measured using all EFOs 
outside a cone of $10^\circ$ in the forward direction. The forward region was excluded because 
of the strong influence of the proton remnant\,\cite{epj:c6:67}.

Finally, the $D^*$ meson was associated with the closest jet  
(with $E_T^{\rm jet} > 9$\,GeV and \mbox{$|\eta^{\rm jet}| < 2.4$}) in $\eta-\phi$ 
space and requiring 
\mbox{$R \left(= \sqrt{(\eta^{\rm jet}-\eta^{D^*})^2 
+ (\phi^{\rm jet}-\phi^{D^*})^2}\right) <$ 0.6}. 

The combined efficiency for all the above requirements  was about 35\%. A clear $D^*$ mass peak 
above a relatively small background is shown in Fig.~\ref{fig:mass}. Subtraction of the 
background of 634\,$\pm$\,30 candidates, estimated from the wrong-charge sample, gave 
1307\,$\pm$\,53 $D^*$ mesons.  The background was subtracted bin-by-bin as a function of the 
measured fragmentation variable and all other subsequent distributions.

\section{Fragmentation variables and kinematic region}
\label{sec-xsec}

In $e^+e^-$ collisions, at leading order (LO), the two produced charm quarks each carry half 
of the available 
centre-of-mass energy, $\sqrt{s}$. The fragmentation variable of a $D^*$ meson can 
therefore be simply related to one of the two produced jets. In $ep$ collisions, the 
definition of the fragmentation variable is not so simple as only a fraction of the 
available centre-of-mass energy contributes to the production of charm quarks in the 
hard scattering process. However, charm quarks produced in the hard scatter form 
final-state jets of which the meson is a constituent. Therefore, the fragmentation 
variable, $z$, is calculated as $z=(E+p_\parallel)^{D^*}/(E+p_\parallel)^{\rm jet}$, 
where $p_\parallel$ is the longitudinal momentum of the $D^*$ meson or of the jet 
relative to the axis of the associated jet of energy, $E^{\rm jet}$, where all quantities 
are given in the laboratory frame. As the jets are reconstructed as massless objects, 
$z$ simplifies to: 

\begin{equation}
z = (E+p_\parallel)^{D^*}/2E^{\rm jet}.
\end{equation}

The analysis was performed in the photoproduction regime with 
\mbox{$130 < W_{\gamma p} < 280$\,GeV} and \mbox{$Q^2<1$\,GeV$^2$}. 
The $D^*$ meson was required to be in the region $|\eta^{D^*}|<1.5$ and 
\mbox{$p_T^{D^*} > 2$\,GeV}. The $D^*$ meson was included in 
the jet-finding procedure and was thereby uniquely associated with one jet only. 
Each jet associated with a $D^*$ was required 
to satisfy $|\eta^{\rm jet}|<2.4$ and \mbox{$E_T^{\rm jet} > 9$\,GeV}.

Cuts on the minimum jet transverse energy and minimum $D^*$ 
transverse momentum will lead to a bias in the $z$ distribution as 
$z \sim p^{D^*}/E^{\rm jet}$. Therefore the minimum jet transverse energy was chosen 
to be as high as possible and the minimum $D^*$ transverse momentum to be as low as 
possible whilst maintaining statistical precision. With the above requirements, the 
$z$ distribution is unbiased above 0.22.

\section{Fragmentation models}
\label{sec-frag}

Various parametrisations of fragmentation functions have been proposed. Those 
considered in this paper are detailed below.

A parametrisation often used to describe the fragmentation of heavy quarks is the 
function from Peterson et al.~\cite{pr:d27:105} which has the form 

\begin{equation}
f(z) \propto \frac{1}{\left[ z(1-1/z-\epsilon /(1-z))^2\right]} \ ,
\label{eq:peterson}
\end{equation}

where $\epsilon$ is a free parameter. 

The function from Kartvelishvili et al.~\cite{pl:b78:615} has the form

\begin{equation}
f(z) \propto z^\alpha (1 - z),
\label{eq:kartvelishvili}
\end{equation}

where $\alpha$ is a free parameter.

Within the framework of the Artru-Mennessier model~\cite{np:b70:93}, Bowler~\cite{zfp:c11:169} 
developed a fragmentation function for heavy quarks of mass, $m_Q$, which has the form

\begin{equation}
f(z) \propto \frac{1}{z^{1+r_Q b m_Q^2}} \left( 1-z \right)^a 
\exp \left( \frac{-b m_\perp^2}{z} \right),
\label{eq:bowler}
\end{equation}

where $a$ and $b$ are free parameters and $r_Q$ is predicted to be unity. The quantity $
m_\perp$ is the transverse mass of the hadron, $m_\perp^2 = m^2 + (p_T^{\rm rel})^2$, 
where $m$ is the hadron's mass and $p_T^{\rm rel}$ the transverse momentum relative to the 
direction of the quark. The additional freedom given by $r_Q$ 
allows a smooth transition to the symmetric Lund form~\cite{zfp:c20:317} 
($r_Q \equiv 0$) used to describe light-quark fragmentation.

\section{Monte Carlo models}
\label{sec-mc}

Monte Carlo (MC) models were used both to calculate 
the acceptance and effects of the detector response and to extract fragmentation 
parameters. The programmes {\sc Herwig 6.1}\cite{cpc:67:465} and 
{\sc Pythia 6.1}\cite{cpc:135:238} which implement LO matrix elements followed by 
parton showers and hadronisation were used to model the final state. Different 
parameter settings were used in the MC models when correcting the data or when 
extracting fragmentation parameters; the settings used when extracting fragmentation 
parameters are given in Section~\ref{sec-results-mc}. The MC used to correct the 
data had the default settings, apart from the following changes: the fraction 
of charged $D$ mesons produced in a vector state was set to 0.6\cite{epj:c44:351}; 
and the excited $D$-meson production rates were set to non-zero 
values\footnote{These changes correspond to the {\sc Pythia} 
parameters: {\tt PARJ(13) = 0.6}, {\tt PARJ(14) = 0.13}, {\tt PARJ(15) = 0.01}, 
{\tt PARJ(16) = 0.03} and {\tt PARJ(17) = 0.13}.}\,\cite{desy-08-093}.

The ZEUS detector response was simulated in detail using a programme based on 
{\sc Geant 3.13}~\cite{tech:cern-dd-ee-84-1}. The {\sc Pythia 6.1} MC programme was 
used with two different fragmentation schemes: the default which is the Lund string 
model\cite{prep:97:31} modified according to Bowler for heavy quarks; and the 
Peterson fragmentation function with $\epsilon = 0.06$ (see Section~\ref{sec-frag}). 
The \mbox{{\sc Herwig 6.1}} MC programme uses a cluster model\cite{np:b238:492} for 
its fragmentation. 

\section{Data correction and systematic uncertainties}
\label{sec-syst}

The data were corrected for acceptance and effects of detector response using a 
bin-by-bin method with the {\sc Pythia} simulation used as the central MC. 
The distribution of the difference in $z$ between hadron and detector levels is 
symmetric, has a mean of zero and a width of 0.06 units.  To maintain both high purity 
and the validity of the bin-by-bin method, a bin width of at least twice this value 
(0.14 units) was chosen.  The rate 
due to $b$ quarks was subtracted using the {\sc Pythia} MC prediction normalised to a 
previous measurement of jet photoproduction~\cite{pr:d70:012008}. Therefore the 
cross section as a function of $z$ is for processes in which an initial-state charm 
quark hadronises to a $D^*$ meson. A detailed 
analysis~\cite{padhi:phd:2004} of the possible sources of systematic uncertainty was 
performed. The sources are:

\begin{itemize}

\item[$\delta_{1}$] the use of an alternative fragmentation model in the {\sc Pythia} MC 
                  simulation (see Section~\ref{sec-mc}). As the {\sc Herwig} MC simulation 
		  gave a poor description of the data, it was not used to correct the data;

\item[$\delta_{2}$] the $b$ fraction subtracted was changed by (a) $+30\%$ and (b) $-30\%$ in 
                  accordance with the level of agreement between data and {\sc Pythia} MC 
		  predictions~\cite{pr:d70:012008} for jet photoproduction;

\item[$\delta_{3}$] the total energy in the jet reconstructed from the CAL EFOs was varied 
                  by (a) $+3\%$ and (b) $-3\%$ in the simulation, in accordance with the 
		  uncertainty in the jet energy scale;

\item[$\delta_{4}$] the range of $W_{\gamma p}$ was changed to (a) \mbox{$124 < W_{\gamma p} < 267$\,GeV}
                  and (b) $136 < W_{\gamma p} < 293$\,GeV, in accordance with the resolution;

\item[$\delta_{5}$] the cut on $E_T^{\rm jet}$ was changed to (a) 10\,GeV and (b) 8\,GeV,
                  in accordance with the resolution;

\item[$\delta_{6}$] the value of the cut on $p_T^{D^*}/E_\perp^{\theta>10^\circ}$ was varied 
                  to (a) 0.08 and (b) 0.12;

\item[$\delta_{7}$] the lower (upper) bound on the normalisation region for the wrong-charge 
                  candidates was changed to (a) 0.152 ((b) 0.163)\,GeV.

\end{itemize}

The cuts on $\eta^{\rm jet}$, $\eta^{D^*}$ and $p_T^{D^*}$ were also varied in accordance 
with their resolution and produced negligible effects. The values of the above 
uncertainties for each bin in the normalised cross section, $(1/\sigma)d\sigma/dz$, are 
given in Table~\ref{tab:syst}. The individual systematic uncertainties were added in quadrature 
separately for the positive 
and negative deviations from the nominal cross-section values to obtain the total systematic 
uncertainties. The systematic uncertainties on the fits of the various fragmentation 
parametrisations to the 
data described in Section~\ref{sec-results} were obtained from fits to the cross section 
for each systematic variation. The resulting variations in a given fragmentation parameter 
were added in quadrature to yield the systematic uncertainty on that parameter.

\section{Results}
\label{sec-results}

\begin{figure}[htb]
\begin{center}
~\epsfig{file=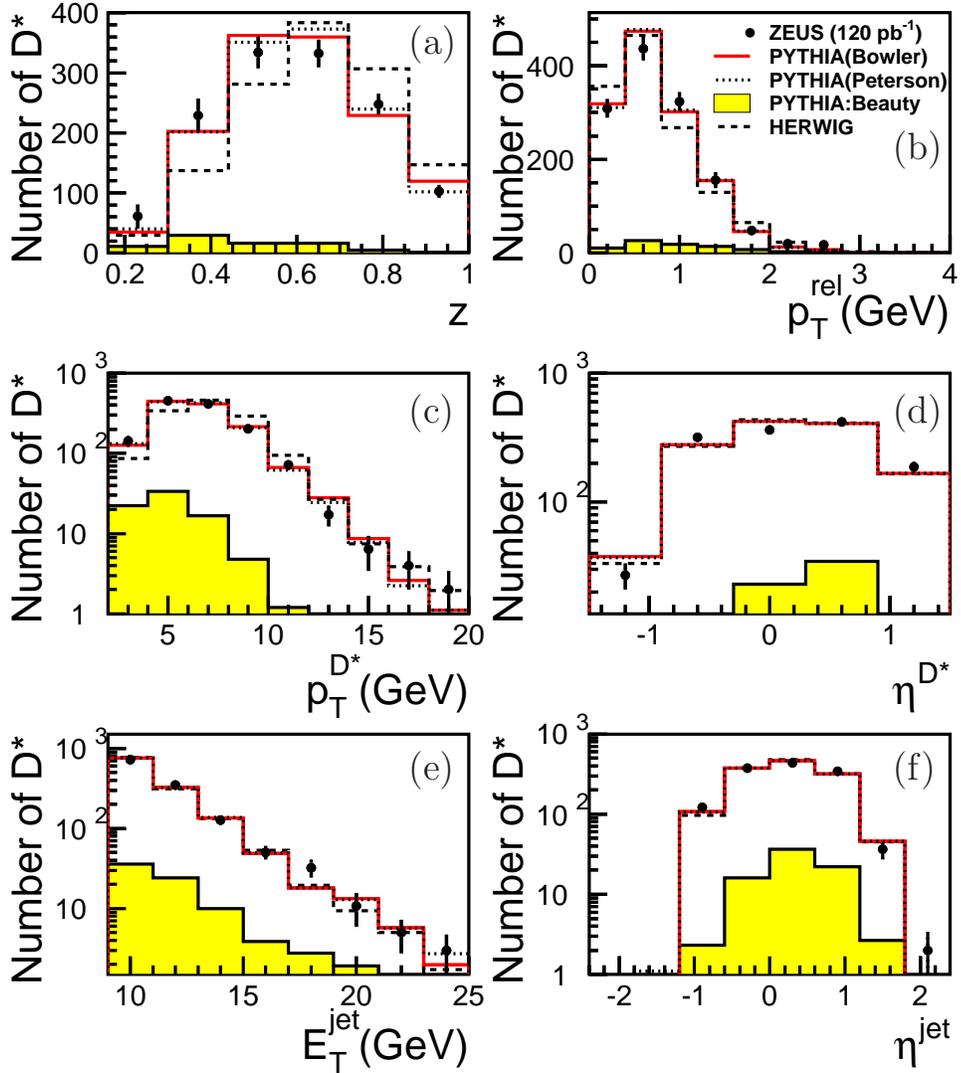,height=16cm}
\put(-210,405){\makebox(0,0)[tl]{\large (a)}}
\put(-27,365){\makebox(0,0)[tl]{\large (b)}}
\put(-210,265){\makebox(0,0)[tl]{\large (c)}}
\put(-27,265){\makebox(0,0)[tl]{\large (d)}}
\put(-210,130){\makebox(0,0)[tl]{\large (e)}}
\put(-27,130){\makebox(0,0)[tl]{\large (f)}}
\end{center}
\caption{Distributions of number of $D^*$ mesons versus (a) $z$, (b) $p_T^{\rm rel}$, 
(c) $p_T^{D^*}$, 
(d) $\eta^{D^*}$, (e) $E_T^{\rm jet}$ and (f) $\eta^{\rm jet}$ for data (points) and MC 
simulations. The data are compared with {\sc Pythia} using the Bowler (solid line) 
and Peterson, with $\epsilon=0.06$, (dotted line) fragmentation functions and with 
{\sc Herwig} (dashed line). The component of beauty production as predicted by 
{\sc Pythia} (shaded histogram) is also shown.}
\label{fig:events}
\end{figure}

The distributions of the variables $z$,  
$p_T^{\rm rel}$, where $p_T^{\rm rel}$ is the transverse momentum of the $D^*$ meson 
relative to the jet, $p_T^{D^*}$, $\eta^{D^*}$, $E_T^{\rm jet}$ and $\eta^{\rm jet}$ 
are shown in Fig.~\ref{fig:events} and compared to the distributions from the MC 
programmes, normalised to the data. Also shown is the prediction of the {\sc Pythia} 
simulation for 
the production of beauty quarks subsequently producing a $D^*$ meson; this amounts to 
about $6\%$. The $z$ distribution is reasonably well described by the {\sc Pythia} MC 
predictions, whereas the {\sc Herwig} prediction does not describe the data.
This can be seen in the differences between the measured $p_T^{D^*}$  distribution and that 
predicted by {\sc Herwig}. The MC predictions for the $E_T^{\rm jet}$ distribution 
are, however, similar and agree reasonably well with the measurement.  For the 
$p_T^{\rm rel}$ distribution, the {\sc Pythia} 
simulations give a good description of the data and are again better than that from 
{\sc Herwig}. This shows that the {\sc Pythia} MC model using both the Bowler and Peterson 
fragmentation for charm quarks gives a good description of the 
transverse as well as the longitudinal component of the $D^*$ fragmentation process. 
The distribution of the pseudorapidities of both jet and $D^*$ are 
similarly well described by both MC programmes. As the {\sc Herwig} MC model is known 
to give a better description than {\sc Pythia} of data~\cite{np:b729:492} sensitive to the 
parton-shower model, the differences shown here suggest that the cluster model 
does not describe the hadronisation process of charm quarks to $D^*$ mesons.

\begin{figure}[htb]
\begin{center}
~\epsfig{file=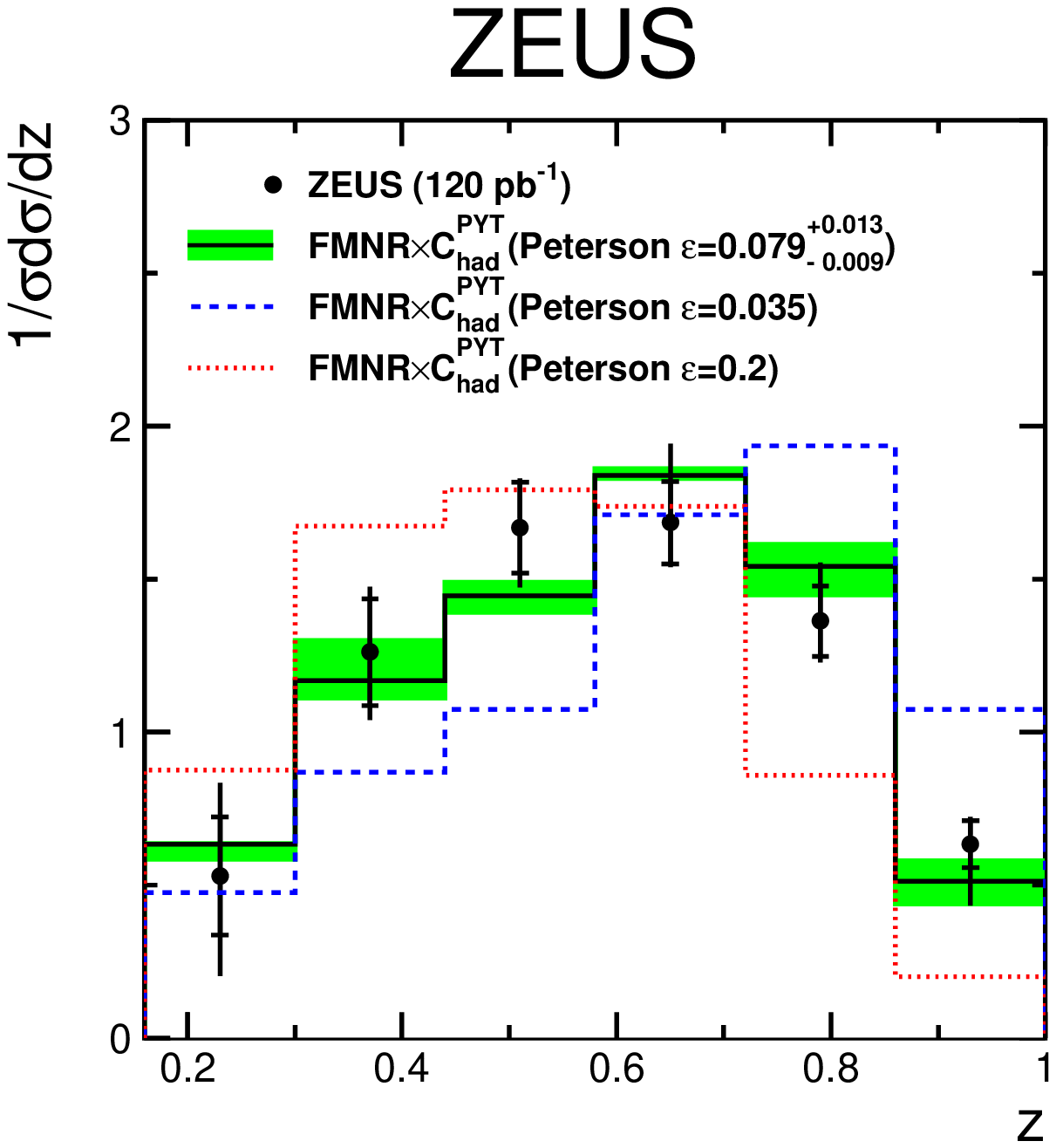,height=9cm}
\put(-33,220){\makebox(0,0)[tl]{\large (a)}}\\
~\epsfig{file=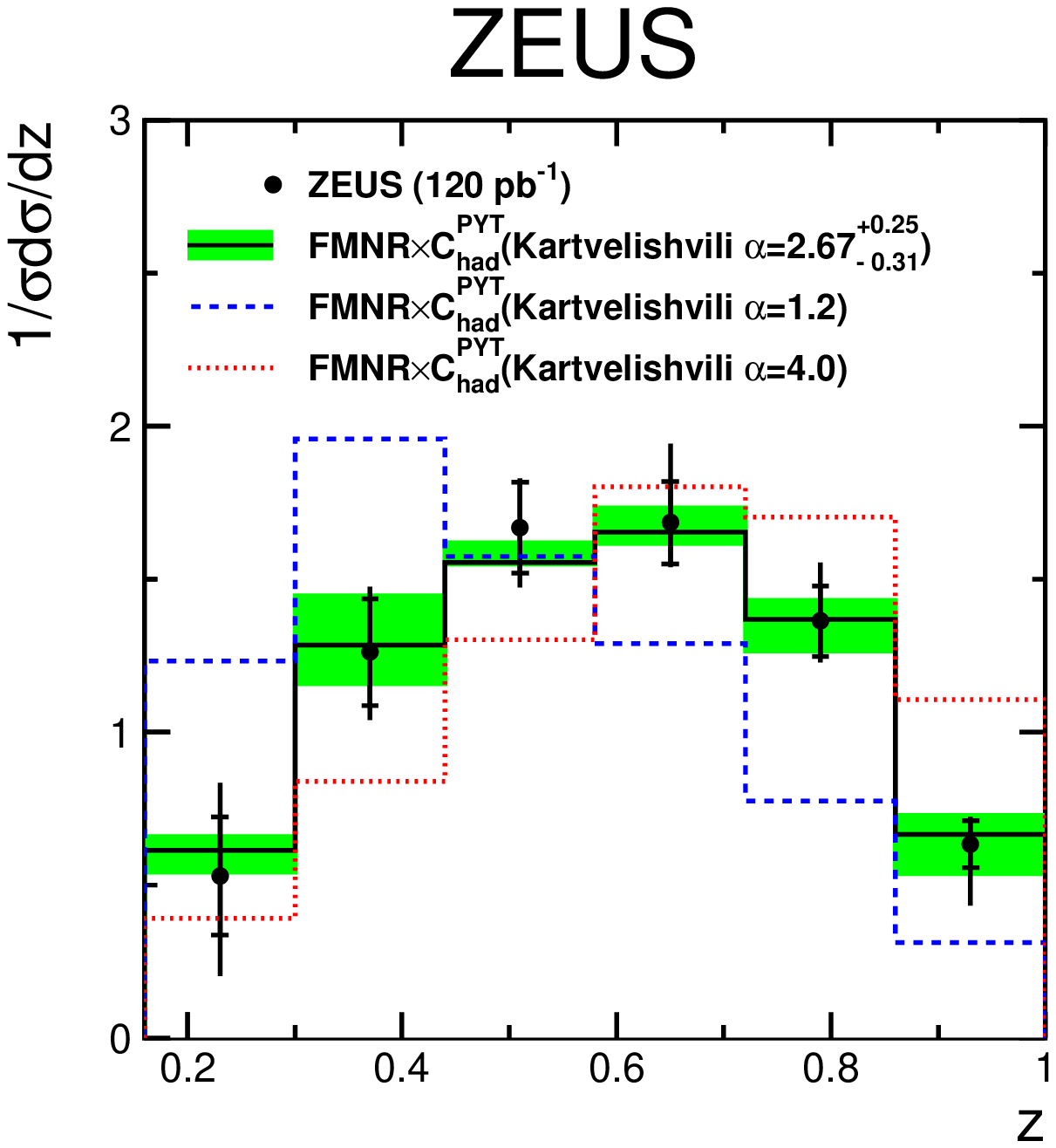,height=9cm}
\put(-33,220){\makebox(0,0)[tl]{\large (b)}}
\end{center}
\caption{Normalised cross section, $1/\sigma(d\sigma/dz)$, for the data (points) compared 
with the predictions of FMNR$\times C_{\rm had}^{\rm PYT}$. (a) the Peterson 
fragmentation function in the calculation is shown with $\epsilon=0.2$ (dotted 
line), $\epsilon=0.035$ (dashed line) and the fitted value 
$\epsilon=0.079^{+0.013}_{-0.009}$\,(stat.$\oplus$syst.) (solid line). (b) the Kartvelishvili 
fragmentation function in the calculation is shown with $\alpha=1.2$ (dashed 
line), $\alpha=4.0$ (dotted line) and the fitted value 
$\alpha=2.67^{+0.25}_{-0.31}$\,(stat.$\oplus$syst.) (solid line). The fitted 
FMNR$\times C_{\rm had}^{\rm PYT}$ predictions are shown with the experimental uncertainties 
of the fit (shaded band).}
\label{fig:z_nlo_old}
\end{figure}

\begin{figure}[htb]
\begin{center}
~\epsfig{file=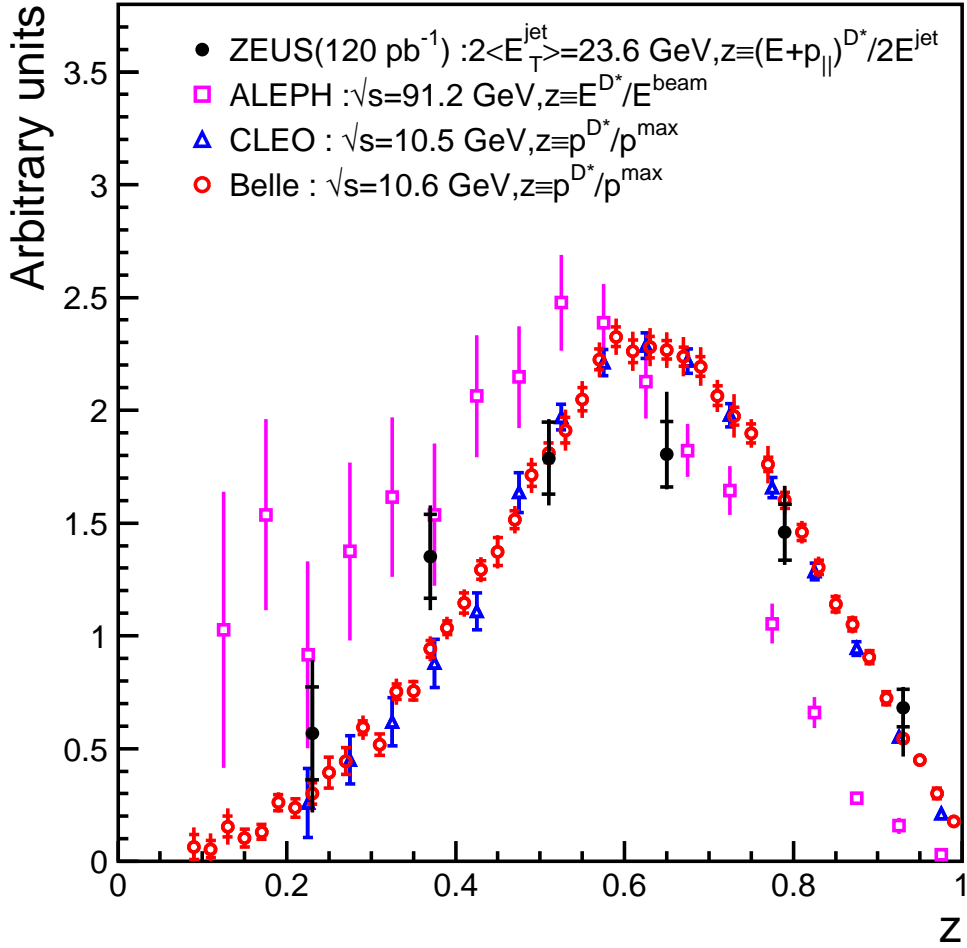,height=14cm}
\end{center}
\caption{$D^*$ fragmentation function for the ZEUS data (solid points) compared to
measurements of the Belle (open circles), CLEO (open triangles) and ALEPH (open squares) 
collaborations in 
$e^+e^-$ collisions. For shape comparison, the data sets were normalised to 
1$/$(bin width) for $z>0.3$. For the ALEPH data, the fragmentation function is measured 
versus the ratio of the energy of the $D^*$ meson and the beam energy, whereas for the 
Belle and CLEO data, the fragmentation function is measured versus the ratio of the momentum 
of the $D^*$ meson and the maximum attainable momentum at the relevant beam energy.}
\label{fig:z_world}
\end{figure}

The normalised differential cross section, $1/\sigma(d\sigma/dz)$, is presented in 
the kinematic region \mbox{$Q^2<1$\,GeV$^2$} and \mbox{$130<W_{\gamma p}<280$\,GeV}, 
requiring at least 
one jet with \mbox{$E_T^{\rm jet}> 9$\,GeV} and $|\eta^{\rm jet}|<2.4$. A $D^*$ meson 
with \mbox{$p_T^{D^*} > 2$\,GeV} and $|\eta^{D^*}|<1.5$ was required to be 
associated with any jet that satisfied the above jet requirements on $E_T^{\rm jet}$ 
and $\eta^{\rm jet}$. The $D^*$ meson was included in the jet-finding 
procedure and was thereby uniquely associated with one jet only. The values of 
the cross section are given in Table~\ref{tab:xsec} and shown in 
Figs.~\ref{fig:z_mc} and~\ref{fig:z_nlo_old} compared to various expectations. In 
Fig.~\ref{fig:z_world}, the same data are shown compared with results from 
$e^+e^-$ experiments.

\subsection{Comparison with fragmentation models in {\sc Pythia}}
\label{sec-results-mc}

\begin{figure}[htb]
\begin{center}
~\epsfig{file=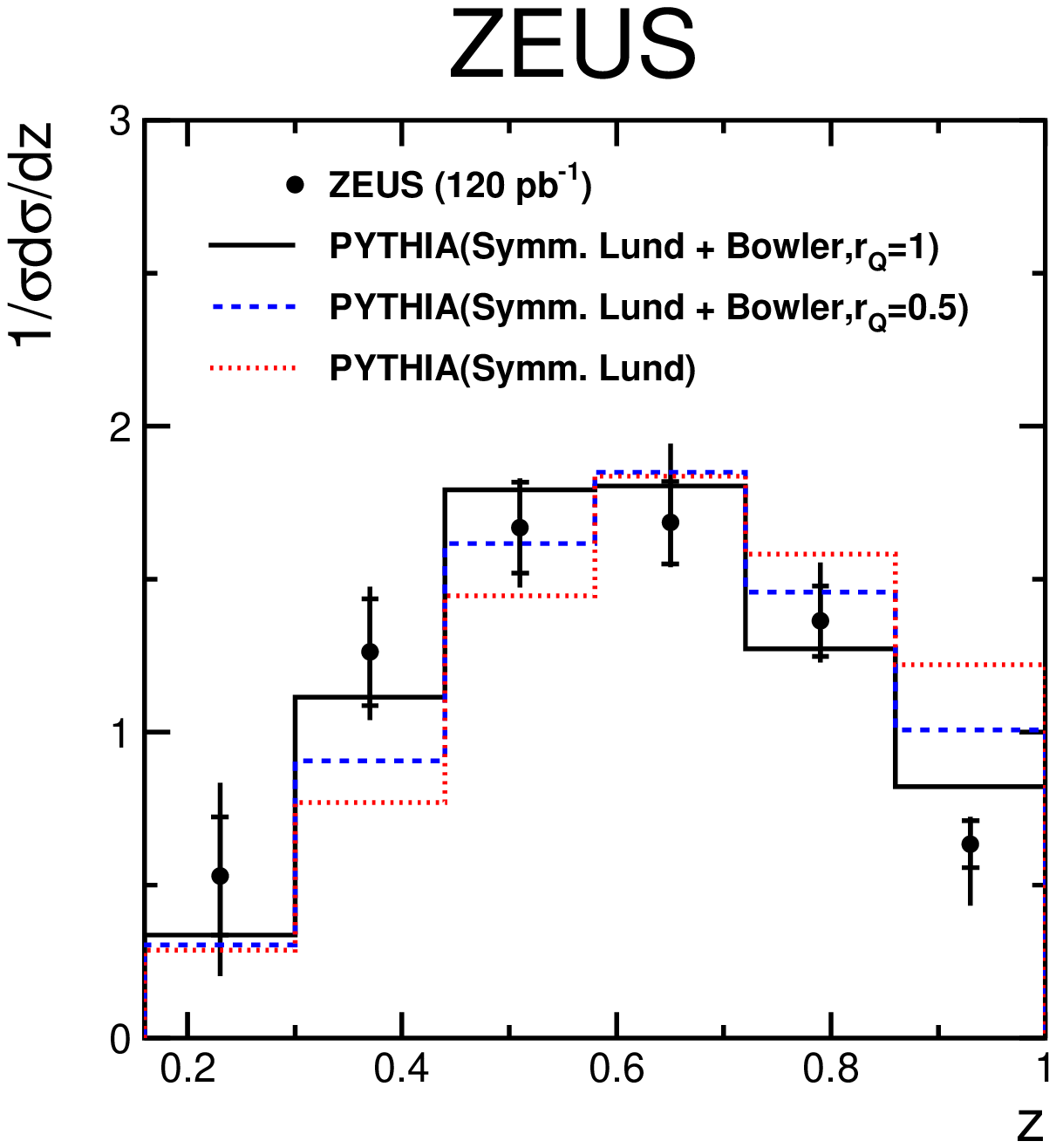,height=9cm}
\put(-33,220){\makebox(0,0)[tl]{\large (a)}}\\
~\epsfig{file=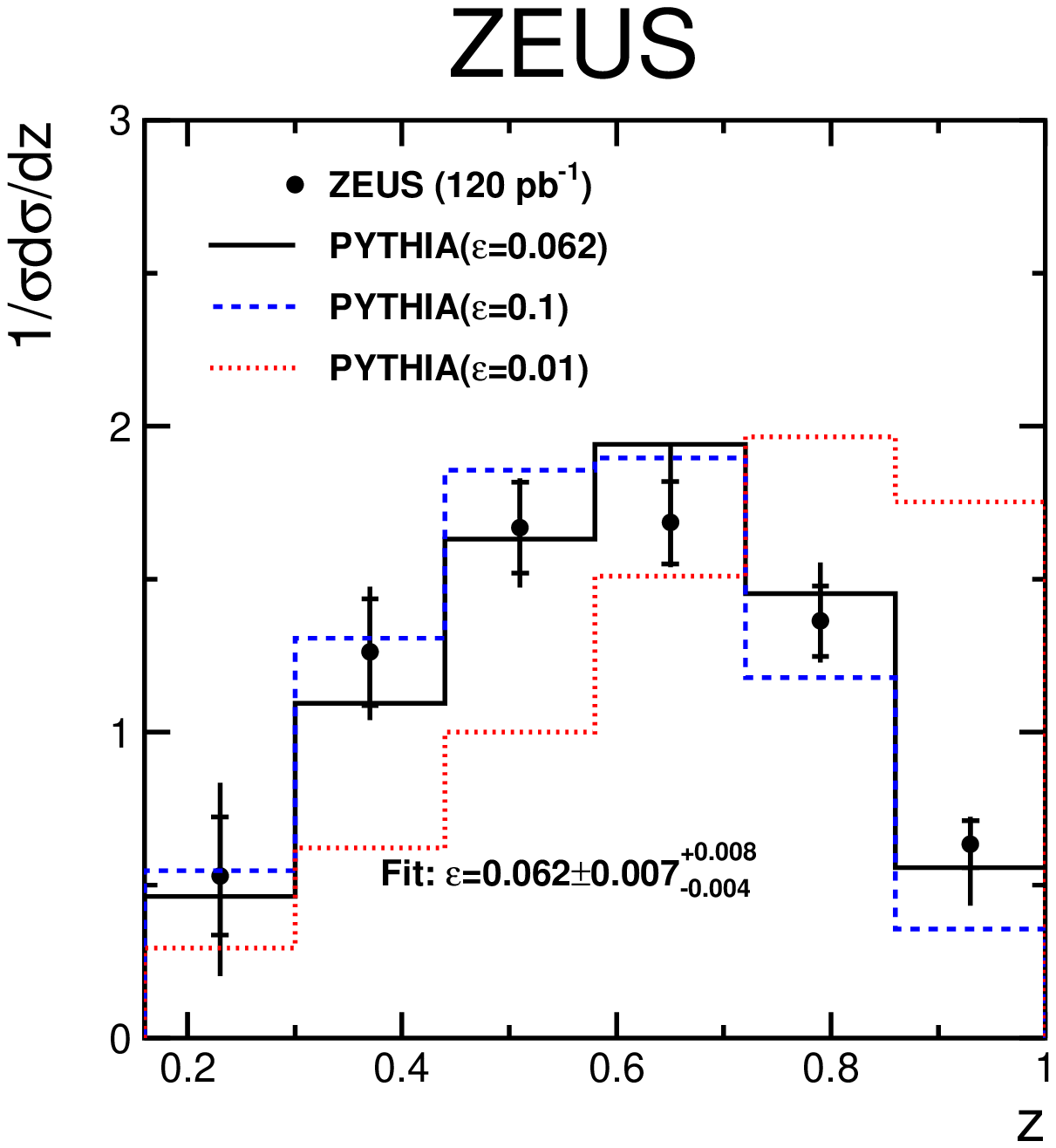,height=9cm}
\put(-33,220){\makebox(0,0)[tl]{\large (b)}}
\end{center}
\caption{Normalised cross section, $1/\sigma(d\sigma/dz)$, for the data (points) compared 
with (a) the symmetric Lund fragmentation modified for heavy quarks (see 
Eq.~\ref{eq:bowler}) with $r_Q=1$ (solid line), $r_Q=0.5$ (dashed line) and the original 
symmetric Lund scheme, $r_Q=0$, (dotted line) as implemented in {\sc Pythia}. The data 
are also compared with (b) the Peterson fragmentation function 
with values of the parameter $\epsilon=0.1$ (dashed line), $\epsilon=0.01$ (dotted 
line) and the fitted value $\epsilon=0.062$ (solid line) as implemented in  
{\sc Pythia}.}
\label{fig:z_mc}
\end{figure}

The normalised cross section is shown in Fig.~\ref{fig:z_mc} compared to the 
{\sc Pythia} MC simulation using different fragmentation models. The original default settings 
for {\sc Pythia 6.1} were used with  
the proton and photon parton density functions set to GRV94 LO~\cite{zfp:c67:433} and 
GRV-LO~\cite{pr:d46:1973}, respectively and a different value for the maximum parton 
virtuality allowed in space-like showers ({\tt PARP(67)} in {\sc Pythia} changed from 
1.0 to 4.0~\cite{rpp:68:2773}). Otherwise, only the fragmentation parameters considered (see 
Section~\ref{sec-frag}) were varied. 

The default fragmentation setting in the simulation is the 
symmetric Lund string fragmentation modified for heavy quarks according to Bowler 
(see Eq.~\ref{eq:bowler}). Three predictions for different values of $r_Q$ are shown 
compared to the data in Fig.~\ref{fig:z_mc}(a). The default prediction with $r_Q=1$ gives a 
reasonable description of the data; as $r_Q$ decreases, the prediction deviates more 
and more from the data.

The Peterson function (see Eq.~\ref{eq:peterson}) and the option to vary 
$\epsilon$ is available within the {\sc Pythia} simulation. The 
value of $\epsilon$ was varied in the range 0.01 to 0.1, with the Lund string 
fragmentation model used for lighter flavours. For each value in the MC simulation, 
the full event record was generated and the kinematic requirements applied, allowing 
a direct comparison to the data. The result of varying $\epsilon$ is shown in 
Fig.~\ref{fig:z_mc}(b). Here it can be seen that values as low as 
$\epsilon=0.01$ are disfavoured, producing a much harder spectrum than the data, 
while values as high as $\epsilon=0.1$ result in too soft a spectrum and are 
therefore also disfavoured. The result of fitting the MC to the data was  
$\epsilon = 0.062 \pm 0.007 ^{+0.008}_{-0.004}$ where the first uncertainty 
is statistical and the second systematic. 
The value is consistent with the default value in the MC of $\epsilon = 0.05$ 
which was obtained from comparisons~\cite{cpc:135:238} with LEP and SLD data at 
the $Z^0$ mass. The fitted value was then used in the MC and the result compared in 
Fig.~\ref{fig:z_mc}(b); the data are well described.

\subsection{Comparison with next-to-leading-order QCD calculations}
\label{sec-results-nlo}

The data were compared with a next-to-leading-order (NLO) QCD 
prediction~\cite{pl:b348:633,*np:b454:3} which is a fixed-order calculation from 
Frixione et al.\ (FMNR). As default, the programme is interfaced to the Peterson 
fragmentation function;  
the function from Kartvelishvili et al. (see Eq.~\ref{eq:kartvelishvili}) was also 
implemented. The other parameters used in the NLO QCD calculation were as follows: the 
renormalisation and factorisation scales were set to 
$\mu = \sqrt{\langle (p_T^c)^2 \rangle + m_c^2}$, where $\langle (p_T^c)^2 \rangle$ 
is the average squared transverse momentum of the two charm quarks and $m_c =$ 1.5\,GeV; 
the proton parton density function was CTEQ5M1~\cite{epj:c12:375}; and the photon parton 
density function was AFG-HO~\cite{zfp:c64:621}.

As the final state particles in the NLO QCD calculation are partons, to enable a 
fair comparison with the data, the predictions were corrected for effects of 
hadronisation using a bin-by-bin procedure according to 
$\Delta \sigma = \Delta \sigma^{\rm NLO} \cdot C_{\rm had}$, where $\Delta \sigma^{\rm NLO}$ is 
the cross section for partons in the final state of the NLO calculation. The 
hadronisation correction factor, $C_{\rm had}$, was defined as the 
ratio of the cross sections after and before the hadronisation process, 
$C_{\rm had}= \Delta \sigma^{\rm Hadrons}_{\rm MC}/\Delta \sigma^{\rm Partons}_{\rm MC}$, where 
the partons used are those after parton showering. The values of $C_{\rm had}$ from 
{\sc Pythia} were used for the central results. As the results of {\sc Herwig} do not 
describe the data (see Section~\ref{sec-mc}), they are used only as a systematic check. 
The prediction from this combination of NLO QCD and hadronisation correction is 
termed ``FMNR$\times C_{\rm had}^{\rm PYT}$''. The values of $C_{\rm had}$ are given 
for {\sc Pythia} and {\sc Herwig} in Table~\ref{tab:xsec}.

The result of varying $\epsilon$ in the Peterson function and $\alpha$ in the 
Kartvelishvili function for the predictions of FMNR$\times C_{\rm had}^{\rm PYT}$ 
are shown in Figs.~\ref{fig:z_nlo_old}(a) and (b), 
respectively. The data again show sensitivity to these fragmentation functions 
and can constrain their free parameters. The results of fits to the data are  
$\epsilon = 0.079 \pm 0.008 ^{+0.010}_{-0.005}$ and 
$\alpha = 2.67 \pm 0.18 ^{+0.17}_{-0.25}$ for the Peterson and Kartvelishvili 
functions, respectively, where the first uncertainty is statistical and the second systematic. 

A number of parameter settings which are commonly used in comparison with 
data\,\cite{np:b729:492} were considered. Using $C_{\rm had}$ from {\sc Herwig} 
gave $\epsilon = 0.094 \pm 0.008$ and $\alpha = 2.46 \pm 0.17$, where the 
uncertainty is statistical only. The effect of the input parameters in the 
NLO QCD programme was checked by changing the renormalisation scale and charm 
mass simultaneously to 2$\mu$ and 1.7\,GeV and 0.5$\mu$ and 1.3\,GeV. The 
different settings gave values of $\epsilon$ ($\alpha$) of 0.082 (2.55) and 
0.077 (2.80), respectively; the uncertainty from the NLO QCD input parameters 
is significantly smaller than the experimental uncertainties.

The default $\epsilon$ value used so far in NLO QCD calculations, extracted from a 
fit~\cite{np:b565:245} to ARGUS~\cite{zfp:c52:353} data, was 0.035. 
As the perturbative part of the production in calculations of $e^+e^-$ 
and $ep$ cross sections depends on the scale of the process and colour connections 
between the outgoing quarks and the proton remnant can have an effect, the values of 
$\epsilon$ extracted with NLO QCD from $e^+e^-$ and $ep$ data may not necessarily be 
the same. This illustrates that care is needed in choosing the appropriate 
fragmentation parameter.

\subsection{Measurement of \boldmath{$\langle z \rangle$} and comparisons with \boldmath{$e^+e^-$} data}
\label{sec-results-ee}

In Fig.~\ref{fig:z_world}, the ZEUS data are shown compared with measurements from 
the Belle~\cite{pr:d73:032002}, CLEO~\cite{pr:d70:112001} and ALEPH\cite{epj:c16:597} 
collaborations in $e^+e^-$ interactions. The Belle and CLEO data are measured at a 
similar centre-of-mass energy of about 10.5\,GeV, whereas the ALEPH data was taken at 
91.2\,GeV. The corresponding scale of the ZEUS data is given by twice the average 
transverse energy of the jet, 23.6\,GeV, and is between the two $e^+e^-$ centre-of-mass 
energies.

Although using a different definition for  $z$, the general features of the data 
presented here are similar to those at $e^+e^-$ experiments. However the ZEUS data 
are shifted somewhat to lower values of $z$ compared to the CLEO and Belle data with the ALEPH 
data even lower. This can be seen more quantitatively by extracting the mean value 
of the distribution, $\langle z \rangle = 0.588 \pm 0.025\,({\rm stat.}) \pm 0.029\,({\rm syst.})$. 
The {\sc Pythia} MC programme was used to extrapolate the phase space to $p_T^{D^*} = 0$ 
and to correct for the subsequent exclusion of the region \mbox{$0 < z < 0.16$}.  
It was also used to correct for the finite bin size. The resulting factor was 
0.961.  The corrected value, 

\begin{equation}
\langle z \rangle = 0.565 \pm 0.024\,({\rm stat.}) \pm 0.028\,({\rm syst.})
\end{equation}

and those from ALEPH, Belle and CLEO are shown in Table~\ref{tab:meanz}. 
It should be noted that the ALEPH data uses the beam 
energy as the scale rather than the jet energy which, due to hard gluon emission, would 
be a better equivalent to the jet energy used in this analysis. The usage of jet energy 
for ALEPH data would lead to an increase in $\langle z \rangle$. Although the 
uncertainties on the current measurement are larger than those from the $e^+e^-$ 
experiments, the value is qualitatively consistent with expectations from scaling 
violations in QCD in which $\langle z \rangle$ decreases with increasing 
energy~\cite{jhep:0604:006}.

\section{Summary}
\label{sec-summary}

The fragmentation function for $D^*$ mesons has been measured in photoproduction 
at HERA using the variable $z=(E+p_\parallel)^{D^*}/2E^{\rm jet}$ and requiring a jet with 
\mbox{$E_T^{\rm jet}> 9$\,GeV} and \mbox{$|\eta^{\rm jet}|<$ 2.4} to be associated 
with a $D^*$ meson in the range \mbox{$p_T^{D^*} > 2$\,GeV} and 
$|\eta^{D^*}|<1.5$. 

The data are compared to different fragmentation models in MC 
simulations and a NLO QCD calculation. The cluster model used in the {\sc Herwig} programme 
does not describe the data. Within the framework of NLO QCD and the 
{\sc Pythia} simulation, the free parameters of the Peterson fragmentation function 
and, for NLO QCD, the Kartvelishvili function have been fitted. 

The value of $\epsilon$ in the Peterson function, extracted within the framework of 
NLO QCD, is different to that extracted using data from $e^+e^-$ collisions. 
As the perturbative aspects of the corresponding calculations and the energy scales are 
different, the results 
are not expected to be the same. Future calculations of charm 
hadron cross sections at NLO QCD at HERA should always use the appropriate values. Within 
the consistent framework given by the {\sc Pythia} model, the extracted fragmentation 
parameters agree with those determined in $e^+e^-$ data.

The fragmentation function and the $\langle z \rangle$ are different to those measured 
at different centre-of-mass energies in $e^+e^-$ collisions; the measured $\langle z \rangle$ is 
higher than the ALEPH data and lower than the CLEO and Belle data, qualitatively consistent 
with the scaling of this variable as predicted by QCD. 

\section*{Acknowledgements}

We thank the DESY Directorate for their strong support and encouragement. The remarkable 
achievements of the HERA machine group were essential for the completion of this work and 
are greatly appreciated. The design, construction and installation of the ZEUS detector 
was made possible by the efforts of many people who are not listed as authors.



{
\def\bibname{\Large\bf References}
\def\refname{\Large\bf References}
\pagestyle{plain}
\ifzeusbst
  \bibliographystyle{./BiBTeX/bst/l4z_default}
\fi
\ifzdrftbst
  \bibliographystyle{./BiBTeX/bst/l4z_draft}
\fi
\ifzbstepj
  \bibliographystyle{./BiBTeX/bst/l4z_epj}
\fi
\ifzbstnp
  \bibliographystyle{./BiBTeX/bst/l4z_np}
\fi
\ifzbstpl
  \bibliographystyle{./BiBTeX/bst/l4z_pl}
\fi
{\raggedright
\bibliography{./BiBTeX/user/syn.bib,%
              ./BiBTeX/bib/l4z_articles.bib,%
              ./BiBTeX/bib/l4z_books.bib,%
              ./BiBTeX/bib/l4z_conferences.bib,%
              ./BiBTeX/bib/l4z_h1.bib,%
              ./BiBTeX/bib/l4z_misc.bib,%
              ./BiBTeX/bib/l4z_old.bib,%
              ./BiBTeX/bib/l4z_preprints.bib,%
              ./BiBTeX/bib/l4z_replaced.bib,%
              ./BiBTeX/bib/l4z_temporary.bib,%
              ./BiBTeX/bib/l4z_zeus.bib}}
}
\vfill\eject

\begin{table}[hbt]
\begin{center}
\begin{tabular}{|c|rrrrrr|}
\hline
 & \multicolumn{6}{c|}{$z$ bin} \\
Source & (0.16, 0.30) & (0.30, 0.44) & (0.44, 0.58) & (0.58, 0.72) & (0.72, 0.86) & (0.86, 1)\\
\hline
$\delta_1$         (\%) & $+$17.0 & $-$1.9 & $-$6.4 & $+$3.2 &  $+$8.8 & $-$21.0  \\
$\delta_{2 \rm a}$ (\%) & $+$8.4  & $+$2.2 & $-$1.0 & $-$1.0 &  $-$1.8 & $-$2.2 \\
$\delta_{2 \rm b}$ (\%) & $+$7.0  & $-$2.2 & $+$0.9 & $+$0.9 &  $+$1.7 & $+$2.0 \\
$\delta_{3 \rm a}$ (\%) & $+$1.4  & $-$5.8 & $+$0.3 & $+$1.2 &  $+$1.6 & $+$3.1 \\
$\delta_{3 \rm b}$ (\%) & $+$4.1  & $+$2.9 & $+$0.4 & $-$0.1 &  $-$3.4 & $-$2.7 \\
$\delta_{4 \rm a}$ (\%) & $-$2.9  & $+$5.3 & $-$0.3 & $+$3.0 &  $-$2.0 & $-$11.0  \\
$\delta_{4 \rm b}$ (\%) & $+$14.0 & $-$1.8 & $-$1.1 & $-$0.5 &  $-$0.8 & $-$2.6 \\
$\delta_{5 \rm a}$ (\%) & $-$6.8  & $+$3.6 & $-$1.9 & $+$2.5 &  $+$5.9 & $-$16.0  \\
$\delta_{5 \rm b}$ (\%) & $-$29.0 & $-$7.9 & $+$2.1 & $+$12.0 & $-$0.6 & $+$4.9 \\
$\delta_{6 \rm a}$ (\%) & $-$39.0 & $+$6.1 & $+$2.7 & $+$2.3 &  $+$2.3 & $+$2.3 \\
$\delta_{6 \rm b}$ (\%) & $+$37.0 & $-$3.3 & $-$3.5 & $-$2.7 &  $-$2.4 & $-$2.4 \\
$\delta_{7 \rm a}$ (\%) & $-$0.3  & $+$0.3 & $+$0.5 & $-$0.5 &  $-$0.3 & $-$0.8 \\
$\delta_{7 \rm b}$ (\%) & $-$1.7  & $+$1.1 & $-$0.1 & $-$0.3 &  $+$0.4 & $-$1.7 \\
\hline
\end{tabular}
\caption{Individual sources of systematic uncertainty (in \%) per bin of the normalised  
cross-section $(1/\sigma)d\sigma/dz$. The description of each variation is given in 
Section~\ref{sec-syst}. 
}
\label{tab:syst}
\end{center}
\end{table}

\begin{table}[hbt]
\begin{center}
\begin{tabular}{|l|ccc||c|c|}
\hline
$z$ bin & $(1/\sigma)d\sigma/dz$ & $\delta_{\rm stat}$ & $\delta_{\rm syst}$ & $C_{\rm had}^{\rm PYT}$ & $C_{\rm had}^{\rm HRW}$ \\
\hline
0.16, 0.30 & 0.53  & $\pm$ 0.19 & $^{+0.23}_{-0.26}$ & 1.82 & 1.43 \\
0.30, 0.44 & 1.26  & $\pm$ 0.17 & $^{+0.12}_{-0.14}$ & 1.58 & 1.08 \\
0.44, 0.58 & 1.67  & $\pm$ 0.15 & $^{+0.06}_{-0.13}$ & 1.28 & 1.00 \\
0.58, 0.72 & 1.68  & $\pm$ 0.14 & $^{+0.22}_{-0.05}$ & 1.18 & 0.91 \\
0.72, 0.86 & 1.36  & $\pm$ 0.12 & $^{+0.15}_{-0.07}$ & 1.02 & 0.85 \\
0.86, 1    & 0.63  & $\pm$ 0.08 & $^{+0.04}_{-0.18}$ & 1.33 & 1.16 \\
\hline
\end{tabular}
\caption{Measured normalised cross-section $(1/\sigma)d\sigma/dz$. The statistical 
($\delta_{\rm stat}$) and systematic ($\delta_{\rm syst}$) uncertainties are shown 
separately. The bin-by-bin corrections for hadronisation (see Section~\ref{sec-results-nlo}) 
are shown for {\sc Pythia}, $C_{\rm had}^{\rm PYT}$, and {\sc Herwig}, 
$C_{\rm had}^{\rm HRW}$.
}
\label{tab:xsec}
\end{center}
\end{table}

\begin{table}[hbt]
\begin{center}
\begin{tabular}{|c|c|c|l|}
\hline
Collaboration & Scale (GeV) & Measured variable & $\langle z \rangle$ $\pm$ stat. $\pm$ syst. \\
\hline
ALEPH & 91.2 & $\langle E^{D^*}/E^{\rm beam} \rangle$ & $0.4878 \pm 0.0046 \pm 0.0061$ \\
Belle & 10.6 & $\langle p^{D^*}/p^{\rm max} \rangle$ & $0.61217 \pm 0.00036 \pm 0.00143$ \\
CLEO  & 10.5 & $\langle p^{D^*}/p^{\rm max} \rangle$ & $0.611 \pm 0.007 \pm 0.004$ \\
ZEUS  & 23.6 & $\langle (E+p_\parallel)^{D^*}/2E^{\rm jet} \rangle$ & $0.565 \pm 0.024 \pm 0.028$ \\
\hline
\end{tabular}
\caption{Mean value, $\langle z \rangle$, of the fragmentation function in $e^+e^-$ collisions, ALEPH, 
Belle and CLEO, compared with the measurement in this paper.  The statistical and systematic 
uncertainties are shown separately.
}
\label{tab:meanz}
\end{center}
\end{table}



%
%
\end{document}